\documentstyle[11pt]{article}
\date{}
\begin{document}
\title{Landau, Abrikosov, Hofstadter:
Magnetic Flux Penetration in a Lattice Superconductor}
\author{David J.E. Callaway*\\
Department of Physics\\
The Rockefeller University\\
1230 York Avenue\\
New York, NY 10021-6399\\
.\\
callaw @ rockvax.BITNET\\
callaway @ physics.rockefeller.edu}
\maketitle

\begin{abstract}
Magnetic flux penetration in superconductors involves a
rich variety of subtle phenomena, much of which is still
poorly understood.  Here these complexities are studied
by formulating the Ginzburg-Landau equations as a lattice
gauge theory.  Their solutions are compared and contrasted
with the (heuristic) Landau model of type I superconductivity,
and the (perturbative) Abrikosov model for type II
superconductors.  Novelties arise as the continuum limit
is approached, related to an effect discovered by
Hofstadter.  Various cautionary remarks pertinent to
large-scale simulations are made.
\end{abstract}

\footnotetext{*Work supported in part by the U.S. Department
of Energy under
Grant No. DOE-AC01-87ER-40325 Task B.}

\newpage
\noindent 1.  \underline{Prolegomena}.

Accurate prediction of magnetic flux penetration
patterns in superconductors poses a formidable
challenge.  The pictures observed are the result
of competition between several configurations
which are degenerate in energy (or very nearly
so).  Intricate structures on many length scales
can result.  Moreover, the mechanics of multivortex
systems have implications for elementary particle
theory and cosmology, so the problem is of fairly
general interest.

The underlying complexity of the situation mandates
a nonperturbative approach.  Here the Ginzburg-Landau
equations are recast as a lattice gauge theory, and
magnetic flux penetration patterns are determined.
Comparisons are made with the classic early models
of Landau (for type I superconductors) and Abrikosov
(type II superconductors).  Discontinuities like
those in the Hofstadter ``butterfly'' pattern arise
while approaching the continuum limit and are
discussed in depth.

\noindent
2.  \underline{Historical background and synopsis}.

Magnetic flux penetration in superconductors has
been studied for over half a century.  It is thus
appropriate to recapitulate some of the history
of the problem as its mathematical structure is
set down.  Most of the results relevant to this
paper stem from the efforts of three authors:
Landau, Abrikosov, and Hofstadter.   Their
contributions are sketched as the problem unfolds.

\noindent a.  Landau

The geometry of the problem considered here is the
same as the one used originally $^{[1,2]}$ by Landau
in 1937.  A square flat plate of superconductor is
situated in the $xy$ plane, and a perpendicular field
${\bf{H}}$ lies in the $\hat{z}$ direction.  The magnetic field
approaches a constant at large $z$.

This early work of Landau predated the Ginzburg-Landau
equations $^{[3]}$ by 13 years.  It missed the essential
difference $^{[4]}$ between type I and type II superconductors
and ignores important effects of flux quantization.
Still, it survives as a textbook $^{[2]}$ model of the
intermediate state in type I superconductors.

Landau predicted $^{[1,2]}$ that the magnetic field must
penetrate a square plate of what is now called a
``type I superconductor'' in a pattern of stripes.  In the above
geometry, the observed pattern in the $xy$ plane is
independent of $x$ and periodic in $y$ (or vice-versa),
requiring a complete spontaneous breakdown of the
\underline{discrete}  $[x {\leftrightarrow{y}}]$ symmetry.
Subsequent experimental work $^{[5]}$ revealed that the
true situation in type I superconductors is far
more baroque, involving patterns on many length
scales.  The patterns do, however, typically possess
a degree of elongation, in \underline{qualitative}
agreement with Landau.  Indeed, if the magnetic field
is applied at an \underline{oblique} angle, the
domain patterns align $^{[6]}$ in a fashion resembling
his model.

A better theoretical understanding of superconductors
follows from the Ginzburg-Landau formalism.  The major
assumption of the approach $^{[3]}$ is that the free
energy density $f(x)$ of a superconductor has an expansion

\begin{equation}
f({\bf x}) = f_{n} + \frac{1}{2}|{\bf D\psi}({\bf x})|^{2}
+ \frac{1}{4}[(|\psi({\bf x})|^{2} - 1)^{2} - 1]\\
+ \frac{1}{2} \kappa^{2} {\bf H}^{2}({\bf x})
\end{equation}

\noindent
where $\psi$ $({\bf x})$ is the order parameter (i.e., the
superconductor wave function), ${\bf H}$$({\bf x})$ is the
magnetic field, ${\bf D = \partial + iA}$ is the covariant
derivative and $f_{n}$ is the free energy of the normal
state.  Units are chosen so as to measure

\begin{eqnarray}
{\bf{x}} \: {\rm{in \: units \: of}} \: \xi \equiv (\Phi_{0}/
2\pi B_{c2})^{\frac{1}{2}} \nonumber \\
{\bf{A}} \: {\rm{in \: units \: of}} \: \xi B_{c2} \nonumber \\
f({\bf{x}}) \: {\rm{in \: units \: of}} \: B^{2}_{c2}/2\pi \kappa^{2} \nonumber
\\
{\bf{H}} ({\bf{x}}) \: {\rm{in \: units \:of}} \:  B_{c2}
\end{eqnarray}

\noindent where $\xi$ is the temperature-dependent coherence length
and $\Phi_{0}$ = $2\pi \hbar c/ e^{*}$\\
$(e^{*} = 2e)$ is the
elementary flux quantum.  Minimization of the free energy
gives its Euler equations of motion:

\newcounter{eqnnum}
\newcounter{eqnletter}
\renewcommand{\theequation}{\arabic{eqnnum}\alph{eqnletter}}
\setcounter{eqnnum}{3}
\setcounter{eqnletter}{1}
\begin{equation}
{\bf D}^{2} \psi + \psi - |\psi|^{2} \psi = 0
\end{equation}

\addtocounter{eqnletter}{1}
\begin{equation}
+ \kappa^{2}[\partial^{2} {\bf A} - \partial
(\partial \cdot {\bf A})] = {\bf J}
\end{equation}

\addtocounter{eqnletter}{1}
\begin{equation}
{\bf J} = Im[\psi^{*} {\bf D}\psi]
\end{equation}

Note the appearance of the Abrikosov parameter $\kappa$.
This Ginzburg-Landau formalism is quite general $^{[7]}$
and can be derived $^{[8]}$ from the BCS theory.

In the present work, the Landau geometry is used exclusively.
The order parameter $\psi$ and vector potential $A$
depend only on $x$ and $y$ in the deep interior of a
large thick superconducting plate.  By use of this
geometry and current conservation, the Euler equations
simplify to

\addtocounter{eqnnum}{1}
\addtocounter{eqnletter}{-2}
\begin{equation}
\partial^{2} r - \frac{J^{2}}{r^3} + r - r^{3} = 0
\end{equation}
\addtocounter{eqnletter}{1}
\begin{equation}
\epsilon_{ab}\partial_{a}(J_{b}/r^{2}) = F
\end{equation}
\addtocounter{eqnletter}{1}
\begin{equation}
J_b = - \kappa^{2}\epsilon_{bc}\partial_{c}F
\end{equation}
\noindent
where $\epsilon_{ab}$ is the familiar Levi-Civita antisymmetric
tensor (note that $J_{a}$ can be eliminated from
these equations).  Here,

\addtocounter{eqnnum}{1}
\addtocounter{eqnletter}{-2}
\begin{equation}
r(x,y) = |\psi(x,y)|^{\frac{1}{2}}
\end{equation}

\addtocounter{eqnletter}{1}
\begin{equation}
F(x,y) = \epsilon_{ab}\partial_{a}A_{b}(x,y)
\end{equation}
(both are real scalars) and all indices are two-dimensional.
The microscopic flux density $B$ is in the $\hat{z}$
direction,
\addtocounter{eqnletter}{1}
\begin{equation}
{\bf B}(x,y) = \hat{e}_z F(x,y)
\end{equation}

\noindent b.  Abrikosov

The parameter $\kappa$ determines whether a superconductor
is type I $(\kappa^{2} < \frac{1}{2})$ or type II
$(\kappa^{2} > \frac{1}{2})$.  At the boundary point
$(\kappa^{2} = \frac{1}{2})$, a partial integral of
Eqs. (4) exists, sometimes called the ``Sarma'' $^{[9]}$
or ``self-dual'' $^{[10]}$ solution:

\renewcommand{\theequation}{\arabic{eqnnum}}
\addtocounter{eqnnum}{1}
\begin{equation}
-\partial^{2} \: ln \: r = 1 - r^{2} = F
\end{equation}

The existence of these two types of superconductor was
first postulated by Abrikosov $^{[4]}$.  He solved the
Ginzburg-Landau equations perturbatively, taking $r$ and
$F$ to be periodic on a rectangle of size $(\Delta x,
\Delta y)$ coherence lengths.  Each rectangle is penetrated
by $\nu$ units of flux

\addtocounter{eqnnum}{1}
\begin{equation}
\Delta x \cdot \Delta y \cdot B \cdot \xi^{2} = \nu \Phi_{0}
\end{equation}

\noindent
where $\nu$ is an integer and $B$ is the $xy$ spatial average
of the microscopic flux.  The Abrikosov solutions are the
leading terms in an expansion in $(B_{c2} - B)$ about the
linear limit, with $B_{c2}$ the critical field above which
the material goes normal (note that $B_{c2} \equiv 1$ here).
To second order in $(1-B)$, the free energy eq. (1) is

\addtocounter{eqnnum}{1}
\begin{equation}
\overline{f(x,y) - f_n} = \frac{1}{2} \kappa^{2}B^2
-\frac{1}{2} \kappa^{2}
(1 - B)^2/[1+(2\kappa^2-1)\beta]+\cdots
\end{equation}

\noindent
where the bar denotes $xy$ spatial average and

\addtocounter{eqnnum}{1}
\begin{eqnarray}
\beta &\equiv& \overline{r^{4}} \: (\overline{r^{2}})^{-2}\nonumber\\
    B &\equiv& \overline{F} \nonumber\\
F(x,y) &=& \overline{F} - \frac {1}{2}\kappa^{2}
[r^{2}(x,y) - \overline{r^2}] + \cdots
\end{eqnarray}
\noindent(compare eqs. [6]).

For type II superconductors $(\kappa^2 > \frac{1}{2})$,
the perturbative free energy eq. (8) is minimized when
$\beta$ is smallest,

\addtocounter{eqnnum}{1}
\begin{equation}
{\rm Type \: II: \beta = \beta_{min} \approx 1.1596}
\end{equation}

\noindent
leading to the prediction of a triangular lattice of
flux tubes $^{[4]}$.  In the type I case, $(\kappa^2
< \frac {1}{2})$, the perturbative formula eq. (8)
predicts $^{[11-14]}$ a complicated series of patterns
with large $\beta$, implying large fluctuations:

\addtocounter{eqnnum}{1}
\begin{equation}
{\rm Type \: I: \beta \approx 1/(1 - 2\kappa^{2})
> 1}
\end{equation}
\noindent
Although these type I patterns are generally elongated, they
differ in detail from Landau's crude model (see 2a).  The
point is that the superconductor flux density $r^{2}$ behaves
like a quantum ``phase space'' distribution $^{[11-12]}$.
The $x$ and $y$ coordinates of the plate are essentially
Fourier conjugates, like position and momentum in quantum
mechanics.  The ``uncertainty principle'' underlying eq. (7)
implies that a flux distribution which is independent of
one coordinate must be sharply localized in the other, rather
than the periodic function envisaged by Landau.  Landau's
model of type I superconductivity is thus (oddly enough)
inconsistent with the Ginzburg-Landau equations.

It is obviously imperative to go beyond the perturbative
formula eq. (8).  Yet a numerical simulation must confront
a novel phenomenon first reported by Hofstadter $^{[15-17]}$
and elucidated in the next section.

\noindent
c. Hofstadter

The Hofstadter phenomenon becomes relevant when the
Ginzburg-Landau equations are formulated on a lattice.
Define complex fields $\psi(m,n)$ and real fields
$F(m,n)$ on sites $(m,n)$ of a lattice $(x = ma,
y = na$; $m,n = 1$, $\cdots, L)$ with lattice spacing $a$.
When the partial derivatives in eqs. (3a) and (4c)
are replaced by covariant differences, the resulting
lattice equations are

\addtocounter{eqnnum}{1}
\begin{eqnarray}
\psi(m+1,n) + \psi(m-1,n) + U(m,n) \psi(m,n+1) \nonumber \\
+ U^{\ast}(m,n-1) \psi(m,n-1) \nonumber \\
= [\epsilon - a^{2} |\psi(m,n)|^{2}]\psi(m,n)
\end{eqnarray}

\renewcommand{\theequation}{\arabic{eqnnum}\alph{eqnletter}}
\addtocounter{eqnnum}{1}
\addtocounter{eqnletter}{-2}
\begin{eqnarray}
\kappa^{2}[F(m+1,n) - F(m,n)] =  J_y(m,n)\nonumber \\
= Im [\psi^{\ast}(m,n) U(m,n)\psi(m,n+1)]
\end{eqnarray}

\addtocounter{eqnletter}{1}
\begin{eqnarray}
\kappa^{2}[F(m,n+1) - F(m,n)]&=& - J_x(m,n) \nonumber \\
&=& - Im[\psi^{\ast}(m,n)\psi(m+1,n)]
\end{eqnarray}

\addtocounter{eqnletter}{1}
\begin{equation}
U(m,n) = U(m-1,n) \times exp[ia^{2}F(m,n)]
\end{equation}

\noindent
where $\epsilon \equiv 4-a^{2}$ and the gauge is chosen
so that the vector potential lies in the $y$ direction.
The lattice spacing $a$ is measured in units of the
continuum coherence length.  Since the desired solutions
fill the $xy$ plane, $F$ and $|\psi|^{2}$ are taken periodic
on a square of size $L \times L$ lattice spacings.  Then the
periodicity of the physical currents $J(m,n)$ implies
that $\psi$ is in general \underline{quasi-periodic}
rather than periodic:

\renewcommand{\theequation}{\arabic{eqnnum}}
\addtocounter{eqnnum}{1}
\begin{eqnarray}
\psi(m,L) = \psi(m,1)\nonumber\\
\psi(L,n) = \psi(1,n)\nonumber\\
\times \: exp[-ia^2 \sum_{m'=1}^L  \sum_{n'=1}^n
F(m',n')]
\end{eqnarray}

\vspace{7mm}
\noindent
Since $\psi(m+L,n+L)$ is single-valued,

\addtocounter{eqnnum}{1}
\begin{equation}
a^{2} \sum_{m=1}^L \sum_{n=1}^L\nonumber\\
F(m,n) = 2\pi p = a^{2}L^{2}B
\end{equation}

\noindent
so $B$ is the flux density per elementary plaquette in
units of the \underline{continuum} $B_{c2}$.  Here the
integer $p$ gives the number of flux quanta penetrating
the $L \times L$ large square.

Equations (12-15), though simple in appearance, imply
a plethora of strange phenomena.  This complexity can
be illustrated by a calculation of the critical magnetic
field $B_{c2}$ on a finite lattice.  When $B$ exceeds this
value, the material goes normal, and the only sensible
solution to these equations is the trivial one where
$\psi(m,n)$ is vanishes.  Near this limit, $\psi(m,n)$
is small, and the nonlinear terms can be neglected.
Then

\addtocounter{eqnnum}{1}
\begin{eqnarray}
F_0(m,n) &=& constant = B = 2\pi p/a^{2}L^{2}\nonumber\\
U_0(m,n) &= & exp[2\pi i \frac{p}{L^{2}}m]
\end{eqnarray}

\noindent
The equation for $\psi(m,n)$ can be simplified by
separating variables, i.e.,

\addtocounter{eqnnum}{1}
\begin{equation}
\psi_{0}(m,n) = \sum_{I=1}^{L}g_{I}(m) \; exp[\frac{-2\pi i In}{L}]
\end{equation}

\noindent
When the nonlinear terms in eq. (12) are dropped, it
becomes
\addtocounter{eqnnum}{1}
\begin{equation}
g_{I}(m+1) + g_{I}(m-1)+2 \: cos[2\pi(IL-pm)/L^{2}]g_{I}(m)
= \epsilon \: g_{I}(m)
\end{equation}

\noindent
which is Harper's equation [18].  The boundary condition
eq. (14) implies that

\renewcommand{\theequation}{\arabic{eqnnum}\alph{eqnletter}}
\addtocounter{eqnnum}{1}
\addtocounter{eqnletter}{-2}
\begin{equation}
g_I(m+L) = g_{I-p}(m)
\end{equation}

\noindent
so that

\addtocounter{eqnletter}{1}
\begin{equation}
g_I(m+L^2/p) = g_I(m)
\end{equation}

\noindent
(it may be necessary to define a ``superlattice'' [15] if
$L^{2}/p$ is nonintegral).  Thus the natural periodicity of the
$g_I(m)$ is

\renewcommand{\theequation}{\arabic{eqnnum}}
\addtocounter{eqnnum}{1}
\begin{equation}
L^{2}/p \equiv 1/\alpha
\end{equation}

\noindent
where $\alpha$ is the (rational) number of flux quanta
per plaquette.

The lattice bulk critical field $B_{c2}(\alpha)$ is
determined from the largest eigenvalue $\epsilon_{max}
(\alpha)$ of eq. (18) via eq. (16):

\addtocounter{eqnnum}{1}
\begin{equation}
B_{c2}(\alpha) = 2\pi\alpha/[4-\epsilon_{max}(\alpha)]
\end{equation}

\noindent
Note that $B_{c2}(\alpha)$ is a function of $\alpha$ alone.

The Hofstadter phenomenon [15] occurs when the eigenvalue
spectrum of eq. (18) is calculated.  The result for
$\epsilon(\alpha)$ is a very striking discontinuous
``butterfly'' pattern, with an intricately organized
hierarchical fine structure.  From eq. 15,

\addtocounter{eqnnum}{1}
\begin{equation}
a^2 = (\frac{2\pi}{B})\alpha
\end{equation}

\noindent
The continuum limit occurs when the lattice spacing
approaches zero at fixed B, implying that the limit
of interest is $\alpha \rightarrow 0$.  But the
expected continuum limit

\addtocounter{eqnnum}{1}
\begin{equation}
\lim_{\alpha\rightarrow0} \: B_{c2}(\alpha) = 1
\end{equation}

\noindent
is \underline{not} obtained smoothly, instead occurring
along the discontinuous upper boundary of the Hofstadter
butterfly.  Needless to say, this is unsettling behavior
for the continuum limit of a lattice gauge theory, for
presumably all thermodynamic functions (and not just
$B_{c2}$) will display rough structure as the continuum
is approached.

Two useful properties which follow from eq. (18) are

\renewcommand{\theequation}{\arabic{eqnnum}\alph{eqnletter}}
\addtocounter{eqnnum}{1}
\addtocounter{eqnletter}{-1}
\begin{equation}
\epsilon_{max}(\alpha) = \epsilon_{max}(1-\alpha)
\end{equation}

\addtocounter{eqnletter}{1}
\begin{equation}
\epsilon_{max}(\alpha+N) = \epsilon_{max}(\alpha)
\end{equation}

\noindent
where $N$ is an arbitrary integer.  Then

\renewcommand{\theequation}{\arabic{eqnnum}}
\addtocounter{eqnnum}{1}
\begin{equation}
B_{c2}(\alpha) = \frac {\alpha}{1-\alpha} B_{c2}(1-\alpha)
\end{equation}

{}From ref. [17] values of $\epsilon_{max}(\alpha)$ can be
extracted (see Table 1).  A plot of $B_{c2}(\alpha)$ versus
$\alpha$ is given in Fig. 1.  Note that away from $\alpha =
1/2$.

\addtocounter{eqnnum}{1}
\begin{eqnarray}
B_{c2}(\alpha) &\approx& 1/(1 - \alpha)\nonumber\\
\epsilon_{max}(\alpha) &\approx& 4 - 2\pi \alpha(1-\alpha)
\end{eqnarray}

\noindent
(though neither is ever a continuous function) and $B_{c2}
(\alpha)$ increases without bound as $\alpha$ approaches one.
When $\alpha$ \underline{equals} one, eqs. (12) have the
trivial solution

\addtocounter{eqnnum}{1}
\begin{eqnarray}
\psi(m,n) &=& 1 \nonumber\\
F(m,n) &=& B
\end{eqnarray}

\noindent
[essentially equivalent to having \underline{no} magnetic
field, viz. eq. (24b), as $B/B_{c2}(1) = 0$].

\noindent
3.  \underline{Numerical solution of the lattice equations}.

\noindent
a.  Method

The lattice Ginzburg-Landau equations, eqs. (12) and (13),
are readily accessible to numerical simulation [19,20],
though it is well to remember the cautionary remarks of
section 2c.  The scheme used here is particularly simple.
First, an initial choice of the $U(m,n)$ and $\psi(m,n)$
is made for a given average $B$.  Then eq. (12) is solved
by relaxation [i.e., each of the $\psi(m,n)$ is determined
from its nearest neighbors].  One hundred sweeps through
the lattice prove sufficient. Given the old $U(m,n)$ and
$\psi(m,n)$, the new $F(m,n)$ are determined from eq. (13)
via a finite Fourier transformation in a single step with
$B$, the average of $F(m,n)$, held fixed.  The new $U(m,n)$
are determined, and the process is repeated.  Typically
about 2,000  loops through the whole algorithm suffice.
The usual checks using different starting conditions were
made.  An easy and adequate initial condition is
$F(m,n) = B;$ $\:$ $\psi(m,n) = 1$.

The limit of interest is that of small $\alpha$, by eq.
(22).  Yet from eq. (20), the natural periodicity of the
system is $1/\alpha$.  Thus, $1/\alpha$ = $L^2/p$ $\geq L$.
The optimal choice for $p$ is therefore $p = L = 1/\alpha$,
and it is used here unless otherwise noted.

\noindent
b.  Vortex arrays in type II superconductors.

One characteristic signature of type II superconductors is
the triangular lattice of flux tubes predicted by Abrikosov.
The $p$ maxima of $F(m,n)$ per $L$ $\times$ $L$ periodic square
are easily seen; their patterns are displayed in fig. 2 for
parameter values $B = 0.9$, $\kappa^2=10$ and various $\alpha$.
Since a triangular lattice involves irrational tangents, it
can never fit exactly on a square lattice; yet the arrays in
fig. 2 form fair approximations to a triangular lattice.  In
fig. 3 the squared distances $d^2$ between lattice points
(taken in units of squared lattice spacing) are plotted
versus $1/\alpha$ along with the Abrikosov value
$d^2$ = $\sqrt{4/3}/\alpha$.  Reasonable agreement is obtained,
though scatter is large.  The lattice spacings are equal
in the $x$ and $y$ directions, so these triangular ``Abrikosov''
lattices are a true nonperturbative prediction (compare [20]).

For these same parameter values the $xy$ average
$\overline{|\psi|^2}$ $\equiv$ $\rho$ is plotted versus
$1/\alpha$ in fig. 4.  As the $\alpha\rightarrow0$ continuum
limit is approached, the value of $\rho$ scatters
discontinuously [as did $B_{c2}(\alpha)$, cf. fig. 1] but
clearly approaches a limit.  Comparison with continuum values
thus requires fairly small $\alpha$ for respectable results.
The values for $\beta(\alpha)$ [cf. eq. (10)] are much better:
e.g. $\beta(1/24) = 1.1596$.

It is well to note the existence of defect structures in
lattice patterns.  Recall from eq. (20) that the natural periodicity
of a pattern is $1/\alpha$ lattice spacings.  If the system
size $L$ is not an integral multiple of $1/\alpha$, defect
structures due to the period mismatch can occur [fig. 5(a)].
Incomplete equilibration can also produce defects [fig.
5(b)], in this case a superimposed triangular lattice of
hexagonal defects.  [Other parameter values are the same as
in fig. 4].  The first problem can be eliminated and the
second reduced by using $p = L = 1/\alpha$.

\noindent
c.  $B$ versus $H$.

The difference between type I and type II superconductors can
be highlighted by comparing $B$, the magnetic flux density
inside the superconductor, with $H$, the applied magnetic field.
Here $B$ is an input parameter given by the spatial average of
$F(m,n)$, while $H$ in the present units is defined by

\addtocounter{eqnnum}{1}
\begin{equation}
H = \frac {1}{\kappa^2} \frac {\partial\overline{f}}{\partial B}
\end{equation}
\noindent
[cf. eq. (8)] and can be calculated with
an elegant virial theorem [21].  It is important
to note that the assumption that $F(m,n)$ is
periodic on an $L \times L$ cell implies constraints by eqs. (13),

\addtocounter{eqnnum}{1}
\begin{equation}
\sum_{m=1}^{L} J_y(m,n) = \sum_{n=1}^{L}J_x(m,n) = 0
\end{equation}

\noindent
For small enough $B$, eqs. (29) are typically violated,
implying that in this limit the only nontrivial solutions to
the Ginzburg-Landau equations involve widely-separated
magnetic vortices.  Thus the plots given in fig. 6 do not
continue to $B = 0$.

In fig. (6a), the type II case $\kappa^2 = 10$ is shown.
Note that, as expected, $H$ is larger than $B$ and
extrapolates to a finite value $H_{c1}$ as $B$ tends to
zero.  At the boundary point $\kappa^2 = \frac {1}{2}$ between
type I and type II superconductors, [fig. (6b)]:

\addtocounter{eqnnum}{1}
\begin{eqnarray}
H(B) &=& B, \: B > B_{c2}\nonumber\\
&=& H_{c2}, B < B_{c2}
\end{eqnarray}

\noindent
while the flux patterns generated are the same as
in the type II limit.

The $B(H)$ curve for type I superconductors is more
subtle.  Figure (6c) displays this function for
$\kappa^2 = 0.35$.  Note that $B(H)$ is double-valued,
with turning points at $H = H_{c2}$ and
$H = H_c > H_{c2}$.  As is well-known, type I
superconductors exhibit ``supercooling,'' leading
to a hysteresis loop in the physically realized
$B(H)$.  Bulk superconductors become normal at
$H = H_c$, while the normal state does not become
superconducting again until $H$ is lowered to
$H = H_{c2} < H_c$.  The curve fig. 6(c) is
metastable for $B < B_{c2}$,

\addtocounter{eqnnum}{1}
\begin{equation}
\frac {\partial^2\overline{f}}{\partial B^2} =
\kappa^2 \frac {\partial H}{\partial B} < 0
\end{equation}

\noindent
Metastability also occurs in the ``effective potential''
of quantum field theory and in the equation of state
of thermodynamic systems (see, e.g., ref. [22]).  There, as
here, the proper behavior of the system can be determined
by a ``Landau construction,'' shown as dotted lines in
fig. 6(c).

The size of the metastable region is determined by the
ratio $H_c/H_{c2}$, which is $1/(\sqrt{2}\kappa)$ in the
continuum.  The metastable region thus becomes more
pronounced as $\kappa$ decreases.  The Ginzburg-Landau formalism
for type I superconductors [11-14] is probably only valid for
$H$ near $H_{c2}$, and true experimental predictions for
equilibrium structures are likely best obtained with
``simulated annealing'' methods such as that applied in
[20] to type II superconductors.  Flux patterns extracted
from the lattice Ginzburg-Landau equations in the metastable
region are generally irregular elongated nucleation lumps
whose periodicity is that of the $L \times L$ system.  Thus
the thickness of the superconducting plate provides an
important scale [11].  Although for reasons discussed above
[in sec. 2(a)] Landau's model of flux penetration in
type I superconductors is incorrect in detail, it may
yet have some qualitative validity.

\vspace{7mm}
\noindent
4.  \underline{Conclusions}.

As should be evident, the study of magnetic flux penetration
in superconductors is a fascinating and difficult subject.
Here the phenomenon was studied by formulating the
Ginzburg-Landau equations as a lattice gauge theory,
following a review of theoretical expectations based upon
heuristic reasoning (Landau) and perturbation theory
(Abrikosov).  Taking the continuum limit of the lattice
gauge theory was a more subtle operation than expected,
involving novelties first discussed in depth by Hofstadter.

Nevertheless, results familiar from continuum theory were
readily obtained for both type I and type II superconductors.
In the latter case the expected triangular ``Abrikosov''
lattice of flux tubes was obtained without the traditional
recourse to perturbation theory.

The ``intermediate state'' of type I superconductors proved
to be most remarkable.  Here the solutions of the classical
Ginzburg-Landau equations of motion were shown to predict a
region of metastability, which perforce limits their domain
of validity.  Flux penetration in type I superconductors seems
to be controlled more by the physics of metastability than by
the Ginzburg-Landau paradigm, though predictions can be made
[11,12].  Time-dependent theoretical [23] and experimental
[24] studies should prove important here, as well as studies
using ``simulated annealing'' methods [20].

Possible practical applications of sensitive metastable
phenomena in the superconducting intermediate state include
a new type of ``dark matter'' detector [11,12] for particle
physics.  It may also be useful to view intermediate state
superconductors as a giant array of Josephson junctions
with dynamic boundaries.

Supercomputer time provided by the Department of Energy proved
useful at several stages of this work.  This manuscript was
typeset in LaTeX by Toni Weil.

\newpage
\begin{center}
{\underline{\Large{Table 1.  $\epsilon_{max}(\alpha)$ for selected $\alpha$.}}}

\vspace{7mm}
\begin{tabular}{cccccl}
\multicolumn{1}{l}{\underline{$\alpha$}} &
\multicolumn{4}{c}{\mbox{}} &
\multicolumn{1}{l}{\underline{$\epsilon_{max}(\alpha)$}} \\
    & &  & & \\
1/2 & {\mbox{}} & {\mbox{}} & {\mbox{}} & {\mbox{}} & $2\sqrt{2}$ \\
    & &  & & \\
1/3 & {\mbox{}} & {\mbox{}} & {\mbox{}} & {\mbox{}} & 1 + $\sqrt{3}$ \\
    & &  & & \\
1/4 & {\mbox{}} & {\mbox{}} & {\mbox{}} & {\mbox{}} & $2\sqrt{2}$ \\
    & &  & & \\
1/5 & {\mbox{}} & {\mbox{}} & {\mbox{}} & {\mbox{}} & 2.96645 \\
    & &  & & \\
1/6 & {\mbox{}} & {\mbox{}} & {\mbox{}} & {\mbox{}} & (5 + $\sqrt{21})^{1/2}$
\\
    & &  & & \\
1/8 & {\mbox{}} & {\mbox{}} & {\mbox{}} & {\mbox{}} &
[6 + (12 + 8 $\sqrt{2})^{1/2}]^{1/2}$
\end{tabular}
\end{center}

\newpage
\begin{enumerate}

\item  L.D. Landau, Phys. Z. Sov. \underline{11} (1937) 129
[JETP \underline{7} (1937) 371]; J. Phys. USSR \underline{7} (1943) 99
[JETP\underline{13} (1943) 377]; and in Collected papers of L.D. Landau,
ed., D. ter Haar (Gordon and Breach, New York, 1967).

\item  L.D. Landau and E.M. Lifshitz,
Electrodynamics of continuous
media (Pergamon, New York, 1984), sec. 57.

\item  V.L. Ginzburg and L.D. Landau,
Zh. Eksp. Teor. Fiz.\underline{20} (1950) 1064.

\item  A.A. Abrikosov, Zh. Eksp. Teor. Fiz.
\underline{32} (1957) 1442 [JETP \underline{5} (1957) 1174].
[An important numerical error in this paper was corrrected by
W.H. Kleiner, L.M. Roth, and S.H. Autler, Phys. Rev.
\underline{133} (1964) A122].  A.A. Abrikosov, Dokl. Acad.
Nauk SSSR \underline{86} (1952) 489.

\item  J.D.Livingston and W. De Sorbo, \underline {in}
Superconductivity, Vol. 2, ed., R.D. Parks, (Marcel
Dekker, New York, 1969); R.P. Huebener, Magnetic flux
structures in superconductors (Springer, New York,
1979); R.N. Goren and M. Tinkham, J. Low. Temp. Phys.
\underline{5} (1971) 465; K.P. Selig and R.P. Huebener,
J. Low. Temp. Phys. \underline{43} (1981) 37; W. Buck,
K.-P. Selig and J. Parisi, J. Low. Temp. Phys.
\underline{45} (1981) 21.

\item  Yu. V. Sharvin, Zh. Eksp. Teor. Fiz. \underline{33}
(1957) 1341 [JETP \underline{6}(1958) 1031]; D.E. Farrell,
R.P. Huebener, and R.T. Kampwirth, J. Low. Temp. Phys.
\underline{19} (1975) 99.

\item  S. Weinberg, Prof. Theor. Suppl. \underline{86}
(1986) 43.

\item  L.P. Gor'kov, Zh. Eksp. Teor. Fiz.
\underline{36} (1959) 1918 [JETP \underline{9}
(1959) 1364].

\item  P.G. de Gennes, Superconductivity of metals and
alloys, (Benjamin, New York, 1966), p. 207;
D. Saint-James, G. Sarma, and E.J. Thomas, Type II
superconductivity, (Pergamon, New York, 1969).

\item  E.B. Bogmol'nyi, Yad. Fiz. \underline{24}
(1976) 861 [Sov. J. Nucl. Phys. \underline{24}
(1976) 4473]; H. de Vega and F. Schaposnik, Phys.
Rev. \underline{D14} (1976) 1100; L. Jacobs and
C. Rebbi, Phys. Rev. \underline{B19} (1978) 4486.

\item  D.J.E. Callaway, Ann. Phys. (N.Y.)
\underline{213} (1992) 166.

\item  D.J.E. Callaway, Nucl. Phys. B344 (1990)
627; Nucl. Phys. B (Proc. Suppl.) \underline{17}
(1990) 270.

\item  G. Lasher, Phys. Rev. \underline{154}
(1967) 345; \underline{140} (1965) A523.

\item  K. Maki, Ann. Phys. (N.Y.) \underline{34}
(1965) 363; J. Pearl, Appl. Phys. Lett.
\underline{5} (1964) 65; A.L. Fetter and P.C.
Hohenberg, Phys. Rev. \underline{159} (1967)
330; R.P. Huebener, Phys. Rep. \underline{C13}
(1974) 143; A. Kiendl, J. Low. Temp. Phys.
\underline{38} (1980) 277.

\item  D.R. Hofstadter, Phys. Rev. \underline
{B14} (1976) 2239; Godel, Escher, Bach: an
eternal golden braid (Vintage, New York, 1989).

\item  M. Ya. Azbel, Zh. Eksp. Teor. Fiz.
\underline{46} (1964) 929; $\:$ A.H. MacDonald
Phys. Rev. \underline{B28} (1983) 6713; $\:$
R. Rammal and J. Bellisard, J. de Phys.
\underline{51} (1990) 2153; $\:$ A. Moroz, Max
Planck preprint MPI-Ph/91-22.

\item  Y. Hasegawa, P. Lederer, T.M. Rice, and
P.B. Wiegmann, Phys. Rev. Lett. \underline{63}
(1989) 907.

\item  P.G. Harper, Proc. Roy. Soc. Lond.
\underline{A68} (1955) 874.

\item  D.J.E. Callaway and R. Petronzio, Nucl.
Phys. \underline{B280} [FS18] (1987), 481;
P.H. Damgaard and U.M. Heller, Phys. Rev. Lett.
\underline{60} (1988) 1246; D.J.E. Callaway and
L.J. Carson, Phys. Rev. \underline{D25} (1982)
531; D.J.E. Callaway and A. Rahman, Phys. Rev.
Lett. \underline{49} (1982) 613.

\item  M.M. Doria, J.E. Gubernatis, and D. Rainer,
Phys. Rev. \underline{B41} (1990) 6335.

\item  M.M. Doria, J.E. Gubernatis, D. Rainer,
Phys. Rev. \underline{B39} (1989) 9573.

\item  D.J.E. Callaway and D.J. Maloof, Phys. Rev.
\underline{D27} (1983) 406; D.J.E. Callaway, Phys.
Rev. \underline{D27} (1983) 2974.

\item  H. Frahm, S. Ullah, and A.T. Dorsey, Phys.
Rev. Lett. \underline{66} (1991) 3067; $\:$ F. Liu,
M. Mondello, and N.D. Goldenfeld, Phys. Rev. Lett.
\underline{66} (1991) 3071.

\item  P.R. Solomon and R.E. Harris, Phys. Rev.
\underline{B3} (1971) 2969.

\end{enumerate}

\newpage
\noindent {\underline{Figure Captions}}

\vspace{.25in}
\noindent {\underline{Figure 1}}

Critical field $B_{c2}(\alpha)$ versus $\alpha$
(points).  The line is the function $1/(1 - \alpha)$,
displayed for comparison.

\vspace{.25in}
\noindent {\underline{Figure 2}}

Unit cells of vortex lattices for $\kappa^2 = 10$ and
$\alpha = 1/8, 1/12, 1/16, 1/20$, and $1/24$;
respectively.
Lines are drawn to guide the eye.

\vspace{.25in}
\noindent {\underline{Figure 3}}

Squared distance (in units of $a^2$) $d^2$ $\:$ between
vortices plotted versus $\alpha$ for regular lattices.
Dots show $d^2$ versus $1/\alpha$ (N.B. $\:$ $\alpha = 10$ and
$13$ correspond to square lattices).  Abrikosov result
$\sqrt{4/3}/\alpha$ shown as straight line.

\vspace{.25in}
\noindent {\underline{Figure 4}}

Average $|\overline{\psi}|^2$ = $\rho$ versus $1/\alpha$
for $\kappa^2 = 10$ and $B = 0.90$.

\vspace{.25in}
\noindent {\underline{Figure 5}}

\noindent
Vortex patterns with defects (shaded).\\
{a) $\alpha = 1/8, L = 12$ $\;$ (periodicity mismatch)}\\
{b) $\alpha = 1/8, L = 16$ $\;$ (insufficient equilibration)}.\\
Lines between vortices are drawn to guide the eye.

\vspace{.25in}
\noindent {\underline{Figure 6}}

\noindent{$B$ versus $H$ for $\alpha = 1/12$} \\
a) $\kappa^2 = 10$  \\
b) $\kappa^2 = 0.5$ \\
c) $\kappa^2 = 0.35$. \\
Solid lines denote the calculated curve; $\:$ dashed lines
the ``Landau construction.''

\end{document}